\documentclass[letterpaper]{article} 
\usepackage{aaai2027}  
\usepackage[hyphens]{url}  
\usepackage{fontawesome5}
\usepackage{graphicx} 
\usepackage{natbib}  
\usepackage{caption} 
\usepackage{amsmath}
\usepackage{amssymb}
\newcommand{\assign}{\,\mathord{\leftarrow}\,}
\usepackage{algorithm}
\usepackage{algorithmic}

\usepackage{newfloat}
\usepackage{listings}
\DeclareCaptionStyle{ruled}{labelfont=normalfont,labelsep=colon,strut=off} 
\floatstyle{ruled}
\newfloat{listing}{tb}{lst}{}
\floatname{listing}{Listing}

\usepackage{booktabs}
\usepackage{multirow}
\usepackage{array}
\newcolumntype{C}[1]{>{\centering\arraybackslash}p{#1}}

\title{TAOT: Topology-Aware Optimal Transport for Dynamic Expert Replica Placement in MoE Training}
\author{
    Lingyun~Zhang\textsuperscript{\rm 1},
    Henghua~Zhang\textsuperscript{\rm 1,\dag},
    Shilei~Gu\textsuperscript{\rm 1},
    Kai~Mo\textsuperscript{\rm 1},
    Shuai~Han\textsuperscript{\rm 1},
    Shiyong~Li\textsuperscript{\rm 1},
    Yanpeng~Wang\textsuperscript{\rm 1},
    Dou~Shen\textsuperscript{\rm 1,\dag}
}
\affiliations{
}

\makeatletter
\newcommand{\arxivtitlefont}{\fontsize{19}{22}\selectfont\bfseries}
\newcommand{\arxivdisplaytitle}{%
  \centering
  TAOT: Topology-Aware Optimal Transport for Dynamic\par
  Expert Replica Placement in MoE Training\par%
}
\newcommand{\arxivauthorfont}{\large\normalfont\itshape\mdseries}
\newcommand{\arxivprojectfont}{\normalsize\normalfont\mdseries}
\def\@maketitle{%
  \def\aaai@theauthors{\if T\showauthors@on\@author\else Anonymous submission\fi}%
  \vbox to \titlebox{%
    \hsize\textwidth%
    \linewidth\hsize%
    \vskip 0.45in minus 0.1in%
    \noindent\begin{minipage}{\textwidth}
      \arxivtitlefont
      \hyphenpenalty=10000
      \exhyphenpenalty=10000
      \arxivdisplaytitle
    \end{minipage}%
    \vskip 0.16in plus 0.05in minus 0.03in%
    \begin{center}
      \begin{minipage}{0.82\textwidth}
        \centering\arxivauthorfont
        \aaai@theauthors\ifhmode\\\fi
      \end{minipage}
    \end{center}%
    \vskip 0.04in plus 0.02in minus 0.01in%
    \begin{center}
      {\arxivprojectfont\faGithub\ \url{https://github.com/baidu-baige/LoongForge}}
    \end{center}%
    \vskip 0.01in plus 0.01in minus 0.005in%
    {\normalsize\aaai@affiliations\ifhmode\\\fi}%
    \vfil%
  }%
}
\makeatother

\nocopyright

\begin{document}

\maketitle
\footnotetext[1]{Baidu, Inc.}
\begingroup
\renewcommand{\thefootnote}{\fnsymbol{footnote}}
\footnotetext[2]{Corresponding authors.}
\endgroup

\begin{abstract}
Mixture-of-Experts (MoE) has become a key architecture for scaling large language models (LLMs), yet its dynamic routing causes severe load imbalance in expert-parallel training. Existing dynamic-replica methods copy hot experts onto idle ranks to share computation, but they optimize load balance alone and ignore the cost of moving expert weights across a multi-node topology, so the resulting cross-node communication can outweigh the balancing gain and inflate training cost. We present TAOT, a topology-aware optimal transport method for dynamic expert-replica placement. TAOT models the overload on hot ranks and the spare capacity on lightly loaded ranks as a balanced entropy-regularized optimal transport problem with a communication-cost matrix, solves it with Sinkhorn-Knopp iterations to produce rank-level flow hints, and combines integer replica matching with token assignment into an executable schedule. At the system level, it overlaps guest-weight transfer with home-expert computation to hide the communication overhead. Experiments show TAOT achieves a $1.43\times$ end-to-end MoE training speedup, reaches balance quality competitive with or better than existing state-of-the-art methods, and attains the lowest weighted expert-communication cost across all configurations, with up to a $74\%$ reduction. 
\end{abstract}

\section{Introduction}

The continued scaling of large language models (LLMs) is pushing training systems from dense Transformers~\citep{vaswani2017attention} toward sparsely activated expert models. Mixture-of-Experts (MoE) replaces the dense feed-forward (FFN) sublayer in a standard Transformer with several feed-forward experts, and each token activates only a few of them, so that model size is partly decoupled from the per-step computation~\citep{shazeer2017outrageously,lepikhin2021gshard,fedus2022switch,cai2025survey}. This property makes MoE a foundational architecture for training trillion-parameter models, and it has been adopted by GLaM~\citep{du2022glam}, Mixtral~\citep{jiang2024mixtral}, DeepSeek-V3~\citep{deepseekv3}, Qwen3~\citep{qwen3}, Kimi-K2~\citep{kimik2}, and Llama 4~\citep{llama4}.

In distributed training, expert parallelism (EP) is the core way to support large-scale MoE~\citep{shoeybi2019megatron,rajbhandari2022deepspeedmoe,hwang2023tutel}. Different experts are placed on different GPU ranks, and after the router selects experts for each token, an All-to-All communication step dispatches tokens to the target ranks for computation. However, as the input, the training stage, and the degree of expert specialization vary, the token count on each expert can be highly skewed~\citep{he2022fastermoe,nguyen2026llep,skiadopoulos2026symi}. In synchronous training, the iteration time is set by the slowest rank, so a rank that holds a few hot experts becomes a global straggler and drags down the overall training throughput.

Prior work mitigates imbalance at two levels: the model-algorithm level and the training-system level. Algorithm-level methods use auxiliary balance losses, capacity factors, expert-choice routing, or auxiliary-loss-free bias adjustment to steer tokens toward a more uniform distribution~\citep{fedus2022switch,deepseekv3,zoph2022stmoe,zhou2022expertchoice}. They improve the long-term statistical distribution, but they must trade off expressiveness, token dropping, wasted capacity, and stability, and they cannot remove the instantaneous imbalance at the micro-batch level. System-level methods keep the routing decision fixed and, once the routing is determined, rearrange where computation happens through parallelism switching, expert remapping, hot-expert replication, token scheduling, or expert-weight migration~\citep{he2022fastermoe,nguyen2026llep,skiadopoulos2026symi,zhai2023smartmoe,zeng2025efficientmoe,zhao2025micromoe,qi2026feplb,liu2026laermoe,megatronecho,lplb}. They do not change the routing target, but they introduce new problems such as replica-weight transfer, optimizer-state migration, extra token forwarding, hardware dependence, or online-solving overhead.

This paper focuses on an underrated but critical question in system-level balancing: where should a replicated expert be placed. A typical replica mechanism reserves a spare slot on a lightly loaded rank, temporarily copies the weights of a hot expert onto it, and lets the lightly loaded rank take over part of the token computation, thereby reducing the straggler load without changing the routing. However, existing placement strategies aim only at ``best load balance'': they treat spare capacity as a homogeneous resource and ignore that the effective bandwidth and cost of intra-node NVLink differ markedly from those of inter-node InfiniBand/RDMA. Two schemes with almost identical balancing quality can therefore incur completely different communication cost, depending on whether expert weights are moved across nodes. This observation leads to the central idea of this paper: MoE replica placement should not optimize load balance alone, but should also treat the communication topology as an objective in the planning. In other words, replica planning must strike the best trade-off between peak-shaving capability and the cost of moving expert weights.

Building on this, we propose TAOT (Topology-Aware Optimal Transport), a topology-aware replica-planning and communication-overlapping scheme for MoE guest-expert placement. While preserving the peak-shaving effect, TAOT steers replication toward intra-node placement as much as possible, and thus greatly reduces cross-node communication and the cost of moving expert weights. Furthermore, once expert communication has been effectively reduced, we hide the weight-distribution overhead introduced by the guest-expert mechanism inside the computation of the home experts on the same rank, achieving computation-communication overlap between hot and cold experts.

We implement TAOT on an in-house training framework built on Megatron-Core. The results show that TAOT improves the end-to-end training speed by $42.82\%$ over Megatron-LM. At the same time, TAOT reaches competitive or state-of-the-art balance quality compared with existing methods at the algorithmic level, and obtains the lowest weighted expert-communication cost across all configurations, with a reduction of up to $74\%$.

The contributions of this paper are summarized as follows:
\begin{itemize}
    \item \textbf{Topology-aware dynamic replica-placement modeling.} TAOT is the first to jointly and explicitly model the per-rank peak-shaving gain and the topology-dependent cost of moving expert weights in guest-expert placement. It obtains the lowest expert-communication cost compared with existing state-of-the-art methods while keeping competitive or state-of-the-art balance results.
    \item \textbf{A GPU-friendly, low-overhead planning algorithm.} We formulate rank-level flow planning as an entropy-regularized optimal transport problem and solve it with Sinkhorn-Knopp iterations, producing a soft topology prior that guides integer replica matching.
    \item \textbf{A communication-overlapping execution design.} We overlap guest-weight transfer with home-expert computation to further hide the communication overhead introduced by the dynamic replicas.
\end{itemize}

\section{Related Work}

MoE was first introduced into deep learning by \citet{shazeer2017outrageously} in a sparsely gated form. GShard and Switch Transformer later scaled it to trillion parameters~\citep{lepikhin2021gshard,fedus2022switch}, and GLaM, Mixtral, and DeepSeek-V3 further confirmed its capacity and performance advantages at a comparable compute budget~\citep{deepseekv3}. To relieve expert load imbalance, routing-level methods typically regulate expert load via auxiliary losses, capacity limits, token dropping, expert choice, or dynamic bias adjustment: ST-MoE targets stability~\citep{zoph2022stmoe}, Expert Choice Routing inverts token-choice so experts pick tokens for better balance~\citep{zhou2022expertchoice}, and BASE Layers and Hash Layers explore alternative assignment schemes~\citep{lewis2021base,roller2021hash}, with surveys likewise listing routing balance as a core issue in MoE scaling~\citep{cai2025survey}. Such methods suppress long-term hot spots at the algorithmic level, but they aim at training stability and expert utilization rather than device-level execution time: an overly strong auxiliary loss weakens semantic routing, capacity limits and expert-choice may drop tokens, and even auxiliary-loss-free routing acts mainly on statistical balance, failing to guarantee equal instantaneous load across EP ranks on any given micro-batch.

System-level dynamic-balancing methods reduce stragglers without changing routing semantics, by adjusting the parallelism strategy, expert layout, expert replicas, or token execution location. One line adaptively tunes parallelism or capacity: FasterMoE replicates hot experts via performance-model-based dynamic shadowing and pipelines the schedule~\citep{he2022fastermoe}, SmartMoE switches online among an offline-built pool of parallel strategies~\citep{zhai2023smartmoe}, and EfficientMoE assigns distinct capacities to hot and cold experts to cut static-graph waste~\citep{zeng2025efficientmoe}. Another centers on hot-expert replication and weight migration: Echo in Megatron-LM quickly matches replicas by overflow amount~\citep{megatronecho}, LLEP greedily migrates overloaded tokens and expert weights to the least-loaded rank as approximate online spill scheduling~\citep{nguyen2026llep}, and SYMI statically shards optimizer states to lower state-migration cost~\citep{skiadopoulos2026symi}. These serve different problems---strategy switching, state migration, token scheduling, or capacity prediction---and none treats topology-aware placement of guest replicas under multi-node EP as a core objective. Additionally, DynamicMoE uses ARIMA to predict load and adjust capacity~\citep{wen2026dynamicmoe}, and FLEX-MoE performs capacity-constrained expert assignment for federated settings~\citep{zhang2026flexmoe}.

\begin{figure*}[t]
\centering
\includegraphics[width=\textwidth]{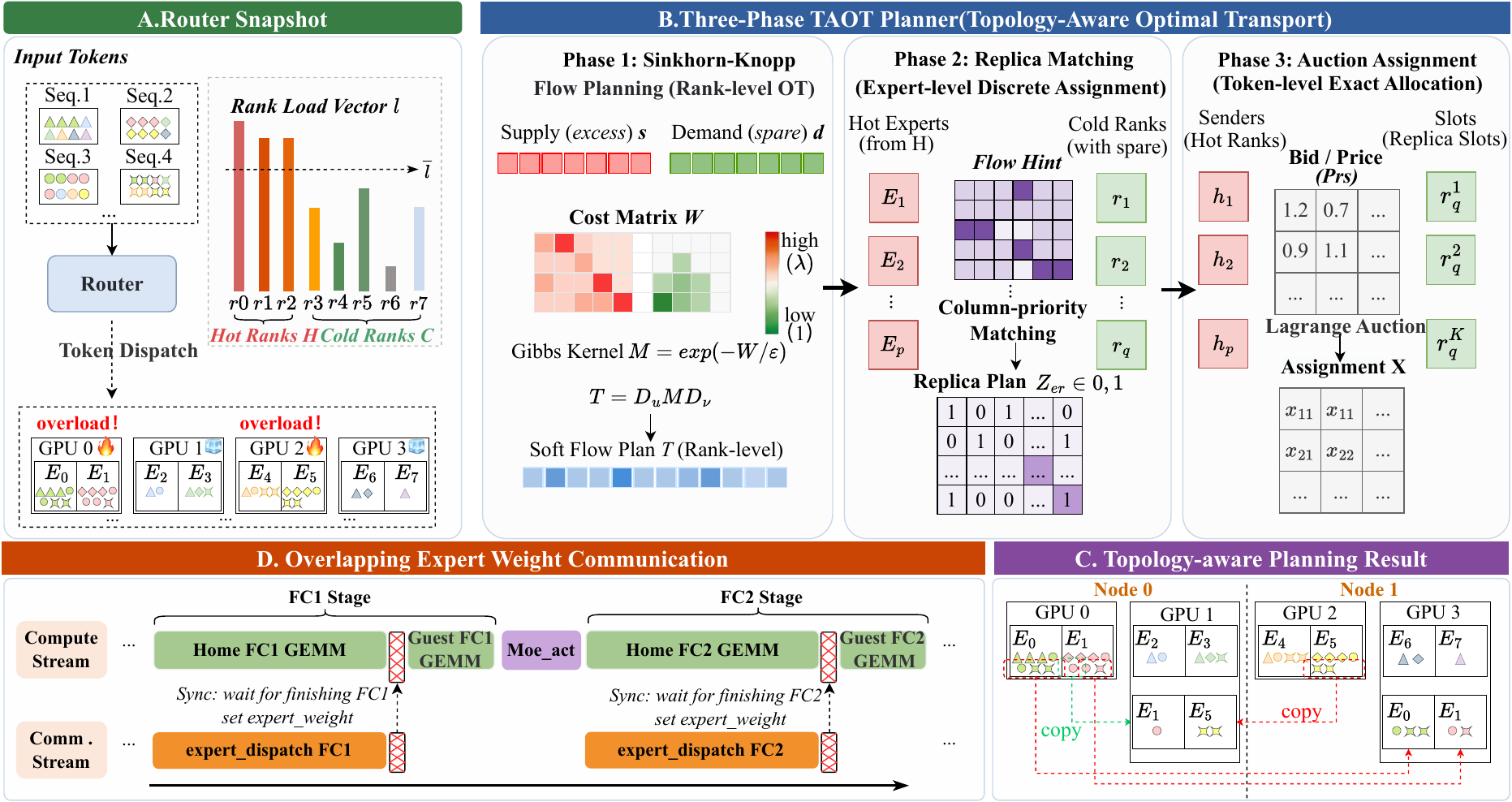}
\caption{The TAOT system architecture.}
\label{fig:overview}
\end{figure*}

Some work also relies on specific communication paradigms or hardware. LAER-MoE splits expert parameters across several devices and, under a fixed parameter- and gradient-communication pattern, plans the recovery location to balance load~\citep{liu2026laermoe}. FEPLB uses the NVLink Copy Engine on Hopper for intra-node movement that barely occupies SM, but confines balancing within a node~\citep{qi2026feplb}. Unlike both, TAOT targets the standard EP setting: it neither depends on LAER-MoE's fixed, complex communication paradigm nor confines balancing within a node as FEPLB does. The closest work to ours, LPLB, models token-to-spare-slot assignment as a linear program and limits replica communication paths with predefined graph structures such as Cube, Hypercube, or Torus~\citep{lplb}, when the EP scale grows or hot spots mismatch the fixed adjacency, local capacity is easily exhausted first while distant idle ranks are constrained by the graph, shrinking the feasible region. TAOT instead uses a continuous communication-cost matrix and a soft topology preference, directly expressing in the optimization the trade-off of preferring intra-node placement while allowing cross-node placement when necessary, which better fits multi-node EP with micro-batch-level dynamic hot spots.

Overall, existing MoE load-balancing research has covered strategy switching, capacity prediction, expert replication, and expert re-layout, but none of the existing schemes considers both the peak-shaving gain and the associated expert-movement cost in the objective function. TAOT effectively fills this gap.

\section{Method}

This section first formalizes the guest-expert placement problem, and then presents a three-stage topology-aware optimal transport (OT) planning algorithm. Figure~\ref{fig:overview} shows the TAOT system architecture. After the router makes its routing decision, we collect the load of each rank and feed it into the three-stage TAOT planner to produce the replica plan. In the forward pass, each rank overlaps the expert-dispatch communication with the home GEMM. In the backward pass, the weight gradients produced by the guest experts are sent back to the home rank through a reverse All-to-All (A2A) and accumulated there.

\subsection{Problem Formulation}

Consider a distributed training cluster with $R$ EP ranks and $E$ experts in total. Each rank holds $E/R$ disjoint experts, and the home rank of expert $e$ is denoted by $r_e$. For a given micro-batch, let $n_{re}$ be the number of tokens that rank $r$ sends to expert $e$. After dispatch, rank $r$ must compute all tokens that its experts $\mathcal{E}_r$ receive from every rank, so its actual computation load is $\ell_{r}=\sum_{e\in\mathcal{E}_r}\sum_{r'}n_{r'e}$, with global mean $\bar{\ell}$. We write the overload and the spare capacity of a rank as $\text{excess}_r=\max(\ell_r-\bar{\ell},0)$ and $\text{spare}_r=\max(\bar{\ell}-\ell_r,0)$, and use them to split ranks into a hot set $\mathcal{H}$ and a cold set $\mathcal{C}$. The load imbalance is defined as

\begin{equation}
\rho = \frac{\max_r \ell_r - \bar{\ell}}{\bar{\ell}}.
\end{equation}

In training, the iteration time is set by the maximum load, so driving $\rho$ close to zero is the core goal. To absorb the imbalance without changing the routing, we reserve $K$ spare slots on each rank to temporarily host the hot experts copied from overloaded ranks, we call these guest experts.

Let $z_{er}\in\{0,1\}$ be the replica-placement indicator ($e$ is held by a hot rank and $r\in\mathcal{C}$ is the target cold rank). Once the placement $\mathbf{z}$ is fixed, the token routing is adjusted accordingly, and $\rho(\mathbf{z})$ is the residual imbalance after adjustment. The replica-placement objective is
\begin{equation}
\begin{aligned}
\min_{\mathbf{z} \in \{0,1\}} & \quad \underbrace{\rho(\mathbf{z})}_{\text{residual imbalance}} + \mu \cdot \underbrace{\sum_{e,\,r} z_{er} \cdot W_{r_e,\,r}}_{\text{weighted replica comm.\ cost}} \\
\text{s.t.} & \quad \sum_{e} z_{er} \leq K, \quad \forall r \in \mathcal{C}
\end{aligned}
\end{equation}
where $\mu>0$ trades off the balance quality against the communication cost, and $W_{r_e,r}$ is the topology communication cost from the home rank of expert $e$ to the target cold rank: it is $1$ for intra-node and $\lambda$ for inter-node.

\subsection{Topology-Aware Optimal Transport Planning}

Solving the two-objective problem jointly is complex, so we decompose it into three stages. Phase~1 plans the global flow direction at the rank level with optimal transport and builds a topology-aware soft flow hint. Phase~2 works at the expert level and combines the flow hint, the spill amount, and the spare capacity to produce an integer replica assignment in a column-first manner. Phase~3 works at the token level and uses a Lagrange auction to finish the exact assignment of each rank to the spare slots.

\textbf{Phase 1: Sinkhorn-Knopp topology-aware flow planning.} Phase~1 sets up a rank-level balanced optimal transport problem with a topology cost: taking the overload $\mathbf{s}$ of each rank as the supply and the spare capacity $\mathbf{d}$ as the demand, it seeks a transport plan under the topology cost matrix $W$:
\begin{equation}
\begin{aligned}
T^* = \arg\min_{T \geq 0} & \quad \langle T, W \rangle \\
\text{s.t.} & \quad T\mathbf{1} = \mathbf{s},\quad T^\top\mathbf{1} = \mathbf{d}.
\end{aligned}
\end{equation}
This OT needs an LP solver and is hard to implement on GPUs. Adding a negative-entropy regularizer relaxes it, and the optimal solution then has a Gibbs-kernel structure $T^*_\varepsilon=\text{diag}(\mathbf{u})M\,\text{diag}(\mathbf{v})$, where $M_{ij}=\exp(-W_{ij}/\varepsilon)$ and the scaling vectors $(\mathbf{u},\mathbf{v})$ are obtained by alternating Sinkhorn-Knopp GEMV iterations. When $\varepsilon=\lambda$, the ratio of the intra-node to the inter-node kernel value is $\exp((\lambda-1)/\lambda)>1$, which forms a soft topology preference. Expanding the rank-level plan to the expert dimension by home rank gives the OT flow-hint matrix $T_{\text{er}}$, whose normalized form serves as the third-level scoring signal in Phase~2.

\textbf{Phase 2: Column-first iterative matching.} Phase~1 gives a continuous flow reference, but replica placement is inherently an integer decision: each cold rank holds at most $K$ complete replicas, and several cold ranks may compete for the same hot expert. Phase~2 therefore refines the continuous flow into the binary decision of ``which expert goes to which spare slot.'' It first sorts the experts within each hot rank by load in ascending order and, through cumulative-excess differencing, computes the spill amount $\text{spill}_e$ that each expert can share out. When expert $e$ is placed on cold rank $r$, this match transfers at most $\min(\text{spill}_e,\text{spare}_r)$ tokens, i.e., the maximum balance improvement of the match. We score each candidate placement of an expert accordingly:
\begin{equation}
\begin{aligned}
\text{score}_{er} = {} & \underbrace{\min(\text{spill}_e,\ \text{spare}_r)}_{\text{main: balance gain}} + \alpha \underbrace{B_{er}}_{\text{second: topology pref.}} \\
& + 0.1\alpha \underbrace{(T_{\text{er}})_{\text{norm}}}_{\text{third: OT flow hint}}
\end{aligned}
\end{equation}
where $\alpha=\bar\ell R/E$ is a scale-alignment factor that makes the main term far larger than the second, ensuring that balance takes priority over topology preference. $B_{er}$ is the topology-preference matrix ($1$ intra-node, $w_{\text{inter}}<1$ inter-node, $0$ for itself), and the OT hint only acts as a global reference when scores tie. We adopt column-first iterative matching from the viewpoint of the cold ranks: in each round every cold rank independently picks the highest-scoring candidate expert, conflicts are resolved by an arbitration mechanism, and losers re-select in the next round (see the Phase~2 part of Algorithm~\ref{alg:taot}).

\textbf{Phase 3: Lagrange auction token assignment.} Phase~2 fixes ``expert $e$ plans $A_{es}$ tokens for spare slot $s$,'' but the tokens of expert $e$ are spread over several ranks, so we still need the actual contribution $X_{r'es}$ of each rank $r'$, subject to $\sum_{r'}X_{r'es}=A_{es}$ and $\sum_s X_{r'es}\leq n_{r'e}$. A direct proportional split would introduce floating-point truncation error and ignore the topology preference. Phase~3 therefore introduces a Lagrange multiplier (price) $p_{r'}$ for the rank-capacity constraint. In each round every spare slot bids for a source rank by net gain $B_{r's}-p_{r'}$ (topology bonus minus current price), after a winning rank is assigned, its price increases monotonically, so its competitiveness naturally decays in the next round. This embeds the topology preference while spreading the load evenly (see the Phase~3 part of Algorithm~\ref{alg:taot}).

\begin{algorithm}[t]
\caption{TAOT Planning (Phase 2 \& Phase 3)}
\label{alg:taot}
\textbf{Input}: $\text{spill}_e$, $\text{spare}_r$, topology-preference matrix $B$, OT hint $T_{\text{er}}$, per-rank holdings $n_{re}$, replica number $K$, price step $\varepsilon$, iterations $T_{\max}$\\
\textbf{Output}: $A_{es}$ (planned amount), $X_{rs}$ (tokens rank $r$ sends to spare slot $s$)
\begin{algorithmic}[1]
\STATE \textit{// Phase 2: column-first iterative matching}
\STATE $U_{er}\assign 0$;\; $A_{es}\assign 0$
\FOR{$k=0,\ldots,K-1$}
    \STATE $m_r \assign 0$
    \FOR{$\text{inner}=0,\ldots,|\mathcal{C}|-1$}
        \STATE $s_{er} \assign \min(\text{spill}_e,\text{spare}_r) + \alpha B_{er} + 0.1\alpha(T_{\text{er}})_{\text{norm}}$
        \STATE $s_{er} \assign -\infty$ if $e\in\mathcal{E}_r \vee U_{er} \vee m_r \vee \text{spare}_r{\leq}0 \vee \text{spill}_e{\leq}0$
        \STATE $e^*_r \assign \arg\max_e s_{er}$;\; $\mathcal{V}\assign\{r: \max_e s_{er}>-\infty\}$
        \STATE $P_{er} \assign \mathbf{1}[r\in\mathcal{V}]\,\mathbf{1}[e^*_r{=}e]$;\; $r^*_e \assign \arg\max_r(s\odot P)_{er}$
        \STATE $\delta_{er} \assign \mathbf{1}[r^*_e{=}r]P_{er}$;\; $g_{er}\assign\delta_{er}\min(\text{spill}_e,\text{spare}_r)$
        \STATE $\text{spill}_e \mathrel{-}{=} \sum_r g_{er}$;\; $\text{spare}_r \mathrel{-}{=} \sum_e g_{er}$
        \STATE $U_{er}\mathrel{|}{=}\delta_{er}$;\; $m_r\mathrel{|}{=}(\sum_e\delta_{er}>0)$
    \ENDFOR
    \STATE $A \assign A + \text{scatter}(g,\ \text{slot }k)$
\ENDFOR
\STATE \textit{// Phase 3: Lagrange auction token assignment}
\STATE $p_r \assign 0$;\; $X_{rs}\assign 0$
\FOR{$t=1,\ldots,T_{\max}$}
    \STATE $s_{rs} \assign B_{rs} - p_r$;\; $s_{rs}\assign-\infty$ if $n_{r,e_s}{\leq}0 \vee A_{e_s,s}{\leq}0$
    \STATE $r^*_s \assign \arg\max_r s_{rs}$;\; $c_{rs}\assign\mathbf{1}[r^*_s{=}r]$
    \STATE $s^*_r\assign\arg\max_s(s\odot c)_{rs}$
    \STATE $x_r \assign \min(n_{r,e_{s^*_r}},A_{e_{s^*_r},s^*_r})\,\mathbf{1}[s^*_r\text{ valid}]$
    \STATE $X_{r,s^*_r}\mathrel{+}{=}x_r$;\; $n_{r,e_{s^*_r}}\mathrel{-}{=}x_r$;\; $A_{e_{s^*_r},s^*_r}\mathrel{-}{=}x_r$
    \STATE $p_r\mathrel{+}{=}\varepsilon\,\mathbf{1}[x_r>0]$
\ENDFOR
\end{algorithmic}
\end{algorithm}
\section{Experiments}

This section evaluates TAOT. We first describe the experimental setup, then give the end-to-end performance comparison, then compare balance quality and communication cost with SOTA methods along two dimensions, then evaluate the scalability, sensitivity, and online planning overhead of TAOT, and finally ablate the roles of the balanced Sinkhorn flow hint and the topology-cost modeling.

\subsection{Experimental Setup}

All end-to-end experiments run on 4$\times$8 A800 GPUs. We use the Qwen3-30B-A3B MoE model on the Pile-test dataset~\cite{gao2020pile}, and compare four methods: Megatron-LM, ECHO, LPLB, and LLEP. As the balancing and communication behavior is corpus-independent, Pile-test is chosen only to provide representative, realistic expert-routing traffic. For metrics, balance quality uses the final imbalance (each rank's largest load deviation from the average after balancing, as a fraction) and the improvement (initial minus final imbalance, in pp). We also report the number of intra-/inter-node expert transfers and the weighted expert-communication cost (with a 1:3 intra/inter ratio).

\subsection{End-to-End Performance}

\begin{figure}[htbp]
\centering
\includegraphics[width=\columnwidth]{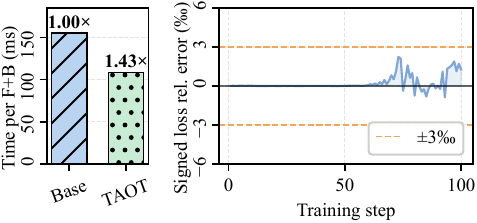}
\caption{TAOT end-to-end speedup and loss consistency.}
\label{fig:end2end}
\end{figure}

Figure~\ref{fig:end2end} shows the end-to-end speedup of TAOT on Qwen3-30B-A3B and the accuracy-consistency check. Performance is measured by the forward-plus-backward (F+B) time of a single iteration (the mean of 10 consecutive steps after 20 warm-up steps). TAOT reduces this time from 155.4\,ms to 108.8\,ms, an end-to-end speedup of $42.82\%$. The right plot compares the lm loss over 100 steps: the error range of TAOT relative to standard EP is $-0.878$\textperthousand{} to $2.237$\textperthousand{}, with a mean absolute relative error of $0.297$\textperthousand{}, staying stably within $\pm 3$\textperthousand{}. This shows that the guest-expert path neither changes the target-expert semantics of the tokens nor breaks the update semantics in which guest gradients are returned and accumulated to the home expert during back-propagation. The speedup comes from better load balance and from expert communication being effectively hidden, not from sacrificing numerical precision.

\subsection{Balance Quality and Communication Cost}

\begin{figure}[htbp]
\centering
\includegraphics[width=\columnwidth]{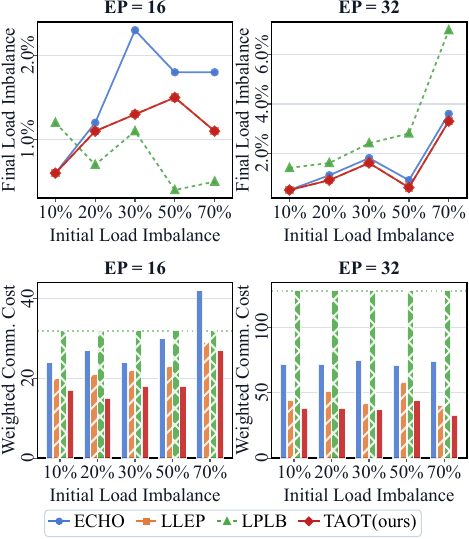}
\caption{Balance effect and weighted expert-communication cost of each method.}
\label{fig:balance}
\end{figure}

To evaluate the balance quality and communication cost of the algorithm itself, we construct five initial imbalance levels of 10\%, 20\%, 30\%, 50\%, and 70\%, and compare with three SOTA methods under EP=16 and EP=32. Table~\ref{tab:balance} summarizes the results of each method, where LPLB uses the Cube topology at EP=16 and the Torus 4$\times$8 topology at EP=32. Figure~\ref{fig:balance} shows how the final imbalance and the weighted cost change with the initial imbalance.

Since LLEP reserves no spare slot, it can reach any target imbalance level by adjusting its transfer threshold. We therefore align LLEP's final imbalance to that of TAOT to compare their routing communication cost fairly under the same balance quality, so LLEP does not enter the balance-quality ranking. In the end-to-end experiments above, expert communication and computation take a ratio of about 1:7. Thus, when two methods differ by only 1\,pp in imbalance, a communication reduction of about 7\% can offset the extra cost from the computation tail. As the EP scale grows, this critical value drops further: the larger the EP, the shorter the per-rank computation window and the more transfer candidates, so communication takes a larger share. We therefore report balance quality and communication cost as two separate metrics, rather than merging them into a single metric via a coefficient.

\begin{table*}[t]
\centering
\setlength{\tabcolsep}{2pt}
\begin{tabular*}{\textwidth}{@{\extracolsep{\fill}}cl|cccc|cccc@{}}
\toprule[1.2pt]
\multirow{2}{*}{\textbf{Initial}} & \multirow{2}{*}{\textbf{Method}} & \multicolumn{4}{c}{\textbf{EP=16}} & \multicolumn{4}{|c}{\textbf{EP=32}} \\
\cmidrule(lr){3-6} \cmidrule(lr){7-10}
 & & \begin{tabular}{@{}c@{}}\textbf{Improve}\\ \textbf{(pp)}\end{tabular} & \begin{tabular}{@{}c@{}}\textbf{Weighted}\\ \textbf{comm.\ cost}\end{tabular} & \begin{tabular}{@{}c@{}}\textbf{Intra-node}\\ \textbf{transfer}\end{tabular} & \begin{tabular}{@{}c@{}}\textbf{Inter-node}\\ \textbf{transfer}\end{tabular} & \begin{tabular}{@{}c@{}}\textbf{Improve}\\ \textbf{(pp)}\end{tabular} & \begin{tabular}{@{}c@{}}\textbf{Weighted}\\ \textbf{comm.\ cost}\end{tabular} & \begin{tabular}{@{}c@{}}\textbf{Intra-node}\\ \textbf{transfer}\end{tabular} & \begin{tabular}{@{}c@{}}\textbf{Inter-node}\\ \textbf{transfer}\end{tabular} \\
\midrule
\multirow{4}{*}{10\%} & ECHO (Megatron-LM) & -9.4 & 24 & 6 & 6 & -9.5 & 72 & 3 & 23 \\
 & LLEP (ICML 2026) & -9.4 & 20 & 5 & 5 & -9.5 & 44 & 5 & 13 \\
 & LPLB (DeepSeek) & -8.8 & 32 & 32 & 0 & -8.6 & 128 & 32 & 32 \\
 & TAOT (Ours) & \textbf{-9.4} & \textbf{17} & 8 & 3 & \textbf{-9.5} & \textbf{38} & 17 & 7 \\
\midrule
\multirow{4}{*}{20\%} & ECHO (Megatron-LM) & -18.8 & 27 & 6 & 7 & -18.9 & 72 & 3 & 23 \\
 & LLEP (ICML 2026) & -18.9 & 21 & 6 & 5 & -19.1 & 51 & 3 & 16 \\
 & LPLB (DeepSeek) & \textbf{-19.3} & 32 & 32 & 0 & -18.4 & 128 & 32 & 32 \\
 & TAOT (Ours) & -18.9 & \textbf{15} & 9 & 2 & \textbf{-19.1} & \textbf{38} & 17 & 7 \\
\midrule
\multirow{4}{*}{30\%} & ECHO (Megatron-LM) & -27.7 & 24 & 9 & 5 & -28.2 & 75 & 6 & 23 \\
 & LLEP (ICML 2026) & -28.7 & 22 & 7 & 5 & -28.4 & 42 & 3 & 13 \\
 & LPLB (DeepSeek) & \textbf{-28.9} & 32 & 32 & 0 & -27.6 & 128 & 32 & 32 \\
 & TAOT (Ours) & -28.7 & \textbf{18} & 9 & 3 & \textbf{-28.4} & \textbf{37} & 16 & 7 \\
\midrule
\multirow{4}{*}{50\%} & ECHO (Megatron-LM) & -48.2 & 30 & 9 & 7 & -49.1 & 71 & 8 & 21 \\
 & LLEP (ICML 2026) & -48.5 & 23 & 8 & 5 & -49.4 & 58 & 4 & 18 \\
 & LPLB (DeepSeek) & \textbf{-49.6} & 32 & 32 & 0 & -47.2 & 128 & 32 & 32 \\
 & TAOT (Ours) & -48.5 & \textbf{18} & 12 & 2 & \textbf{-49.4} & \textbf{44} & 14 & 10 \\
\midrule
\multirow{4}{*}{70\%} & ECHO (Megatron-LM) & -68.2 & 42 & 6 & 12 & -66.4 & 74 & 5 & 23 \\
 & LLEP (ICML 2026) & -68.9 & 29 & 5 & 8 & -66.7 & 40 & 4 & 12 \\
 & LPLB (DeepSeek) & \textbf{-69.5} & 32 & 32 & 0 & -63.0 & 128 & 32 & 32 \\
 & TAOT (Ours) & -68.9 & \textbf{27} & 9 & 6 & \textbf{-66.7} & \textbf{33} & 18 & 5 \\
\bottomrule[1.2pt]
\end{tabular*}
\caption{Comprehensive comparison of load-balancing performance across different EP scales and initial balance conditions. For the improvement and the weighted comm.\ cost, the best in each group is shown in bold.}
\label{tab:balance}
\end{table*}

\textbf{Balance quality.} At EP=16, TAOT is close to the best overall: it ties for best at 10\%, and from 20\% to 70\% it trails LPLB by only about 1\,pp, while turning this small gap into a communication drop that is more valuable to the end-to-end result. This is a ``low-communication trade-off under near-best balance,'' not a plain ranking loss. At EP=32, TAOT is best or tied for best in all cases, showing that as the scale grows it keeps low communication and gains stronger peak-shaving.

LPLB degrades from EP=16 to EP=32 because of the coupling between the hard topology and the load distribution. At EP=16, each rank holds 16 experts, the Cube topology fits the 2-node, 8-card layout, and hot spot overflow can be fully absorbed by nearby ranks within a node. At EP=32, each rank holds only 8 experts, hot spots are split across more nodes, and the Torus 4$\times$8 fixes the migration paths onto adjacent edges, so as the imbalance grows, the capacity around the hot spots is exhausted first while distant idle ranks are unreachable under the topology constraint. TAOT, in contrast, forms a soft topology preference through the balanced Sinkhorn flow hint: it takes the low-cost path when intra-node capacity is enough and accesses cross-node capacity in increasing order of cost when it is not, so the many more low-load ranks at EP=32 instead enlarge its candidate space for offloading.

\textbf{Expert communication cost.} TAOT has the lowest cost across all ten configurations. At EP=16 it is 15--27, up to 53\% lower than LPLB's fixed 32; at EP=32 it is 33--44, up to 55\% lower than ECHO and up to 74\% lower than LPLB.

LPLB plans all spare slots with a predefined Cube/Torus graph, so its cost is set by the topology scale and barely changes with hot spot strength (EP=16: a fixed 32 intra-node and 0 inter-node copies; EP=32: a fixed 32+32 copies, i.e., 128 weighted units), offsetting part of the balance gain end to end. The LLEP result shows the same target imbalance does not imply the same communication efficiency: at 70\% and EP=32, TAOT costs 33 while LLEP costs 40. The difference lies in the placement strategy: LLEP fills idle ranks by load via LPT scheduling and lacks continuous topology-cost modeling, whereas TAOT puts intra-/inter-node communication cost directly into placement and consumes low-cost capacity first, making fewer cross-node copies under the same target.

In summary, at EP=16 TAOT trades a tiny residual imbalance for the lowest communication cost, and at EP=32 it leads in both balance quality and communication cost. This is because a larger scale offers more low-load ranks for offloading, and the balanced Sinkhorn flow hint can make fuller use of the global capacity without being bound by fixed adjacency. This trend is confirmed in Figure~\ref{fig:balance}. In terms of imbalance, TAOT is stably second-best at EP=16 and turns to best or tied for best at EP=32, while LPLB rises to 7.0\% in the high-imbalance region at EP=32, reflecting the shrinking feasible region of the Torus, and ECHO also falls behind in the high-imbalance region for lacking global topology modeling. In terms of communication cost, LPLB is fixed at 32 and 128 and decoupled from the hot spot strength, while TAOT, ECHO, and LLEP transfer on demand, with TAOT the lowest throughout.

\subsection{Scalability and Parameter Sensitivity}

\begin{figure*}[t]
\centering
\includegraphics[width=\textwidth]{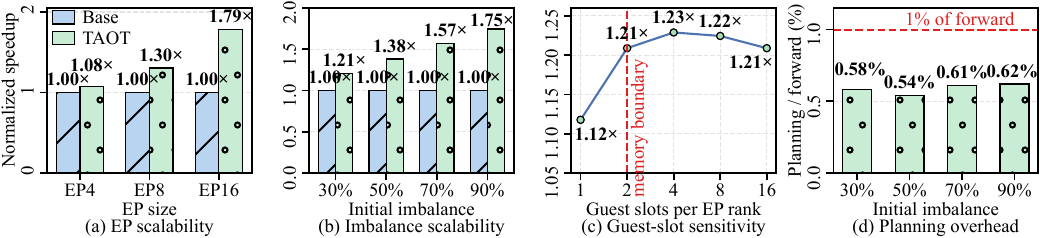}
\caption{Scalability and parameter-sensitivity results of TAOT.}
\label{fig:ablation}
\end{figure*}

This section evaluates TAOT under different EP scales, initial imbalance levels, and numbers of guest slots per rank, as well as the running time of the balancing algorithm under different initial imbalance levels. The imbalance is constructed following LLEP, i.e., routing 30\%, 50\%, 70\%, and 90\% of the tokens to the same rank and sending the rest at random. Figure~\ref{fig:ablation} shows the four groups of results.

\textbf{EP scalability.} As shown in Figure~\ref{fig:ablation}(a), the speedup of TAOT keeps rising as the EP scale grows from EP4 to EP16, up to 1.79$\times$. This shows that a larger EP scale offers richer global spare capacity and a larger search space for topology-aware planning, so TAOT can find more low-communication-cost offloading paths for hot spot tokens.

\textbf{Imbalance scalability.} As shown in Figure~\ref{fig:ablation}(b), when the initial imbalance rises from 30\% to 90\%, the speedup of TAOT grows from 1.21$\times$ to 1.75$\times$, showing stronger adaptability to highly imbalanced cases. The more severe the initial imbalance, the more prominent the computation tail of the hot ranks, and the larger the tail load that the guest-expert mechanism can cut, while the topology-aware cost matrix suppresses cross-node transfer cost, so the peak-shaving gain is kept in the end-to-end execution.

\textbf{Guest-slot sensitivity.} As shown in Figure~\ref{fig:ablation}(c), the gain rises as the number of guest slots per rank grows from 1 to 4 and peaks at 4, then stops growing and drops slightly at 8 and 16, reflecting the trade-off between offloadable capacity and system overhead. Considering both performance and memory (memory stays basically the same as Base when the number of slots is at most 2), 2 guest slots per rank is the better choice in this experiment.

\textbf{Balancing-algorithm overhead.} As shown in Figure~\ref{fig:ablation}(d), the online planning overhead of TAOT stays below 1\% of the forward time and does not scale linearly with the imbalance. This shows that the TAOT algorithm is light enough to be applied at the micro-batch granularity without becoming a new system bottleneck.

In summary, the gain of TAOT grows with the EP scale, thanks to the global spare capacity offered by a larger parallel domain, and also grows with the imbalance, thanks to the effective conversion of the hot spot tail load. The guest slots have a clear empirical optimal range that balances load, communication, and memory under limited resources. Together with the below-1\% online overhead, this shows that the main gain comes from the algorithm design and the system execution path itself.

\subsection{Ablation Study}

To analyze how each design in the TAOT algorithm affects the result, this section performs an algorithm-level ablation at EP=32 and an initial imbalance of 70\%, and reports the mean and standard deviation over several random seeds. Table~\ref{tab:ablation} treats the Phase~1 flow hint and the Phase~2 communication-cost modeling as two switches. Phase~3 only assigns tokens and does not change the replica location, so it is not ablated.

\begin{table}[htbp]
\centering
\small
\setlength{\tabcolsep}{1.5pt}
\begin{tabular}{c c c c c c}
\toprule[1.2pt]
\begin{tabular}{@{}c@{}}\textbf{Phase 1}\\ \textbf{flow hint}\end{tabular} & \begin{tabular}{@{}c@{}}\textbf{Phase 2}\\ \textbf{comm.\ cost}\end{tabular} & \begin{tabular}{@{}c@{}}\textbf{Final}\\ \textbf{imbalance}\end{tabular} & \begin{tabular}{@{}c@{}}\textbf{Intra}\\ \textbf{transfers}\end{tabular} & \begin{tabular}{@{}c@{}}\textbf{Inter}\\ \textbf{transfers}\end{tabular} & \begin{tabular}{@{}c@{}}\textbf{Weighted}\\ \textbf{cost}\end{tabular} \\
\midrule
\multirow{2}{*}{$\times$} & \multirow{2}{*}{$\times$} & $1.95\%$ & $4.67$ & $18.33$ & $59.67$ \\
 & & $\pm 0.18$ & $\pm 1.70$ & $\pm 1.25$ & $\pm 2.36$ \\
\midrule
\multirow{2}{*}{$\times$} & \multirow{2}{*}{$\checkmark$} & $2.00\%$ & $14.67$ & $10.00$ & $44.67$ \\
 & & $\pm 0.82$ & $\pm 0.94$ & $\pm 1.41$ & $\pm 3.86$ \\
\midrule
\multirow{2}{*}{$\checkmark$} & \multirow{2}{*}{$\checkmark$} & $\mathbf{1.48\%}$ & $\mathbf{15.00}$ & $\mathbf{9.67}$ & $\mathbf{44.00}$ \\
 & & $\mathbf{\pm 0.26}$ & $\mathbf{\pm 0.82}$ & $\mathbf{\pm 1.70}$ & $\mathbf{\pm 4.55}$ \\
\bottomrule[1.2pt]
\end{tabular}
\caption{Ablation of the TAOT planning algorithm.}
\label{tab:ablation}
\end{table}

The ablation shows that the two components each do their own job and neither can be dropped. After adding Phase~2, the number of inter-node transfers drops from 18.33 to 10.00 and the weighted communication cost drops from 59.67 to 44.67, while the imbalance stays at about 2\%, showing that it constrains replicas to the intra-node side first while keeping the peak-shaving ability. Adding Phase~1 on top of this further drops the imbalance from 2.00\% to 1.48\%, and the communication cost drops as well, showing that the rank-level flow hint, as a global reference, avoids the case where the local greediness of column-first matching makes some communication paths unreasonable, thus further improving balance quality and lowering communication cost. Under the two working together, the best balance-communication trade-off is reached.

\section{Conclusion}

We present TAOT, a topology-aware optimal transport method for dynamic expert-replica placement, which addresses the load imbalance caused by dynamic routing in MoE expert-parallel training while controlling the cost of moving expert weights. The method unifies the peak-shaving gain and the cross-node communication cost between overloaded and lightly loaded ranks into an entropy-regularized optimal transport problem, and combines Sinkhorn flow hints, integer replica matching, and token assignment to generate an executable schedule. At the system level, it further hides the communication overhead by overlapping guest-weight transfer with home-expert computation.

\bibliography{aaai2027}

\clearpage
\appendix

\section{A\quad Supplementary Details for the Method}

Due to space limits, the Method section of the main paper gives only the key steps of the problem formulation and the three-phase planning algorithm. This appendix provides the complete version: the full problem formulation with all definitions (\S A.1), and the three planning phases with their full derivations and the two planning algorithms in full (\S A.2--A.4).

\subsection{A.1\quad Problem Formulation}

To cast the guest expert placement problem into a unified optimization framework rather than relying entirely on topology-unaware greedy heuristics, we first establish a mathematical model.

Consider a distributed training cluster with $R$ EP ranks deploying $E$ experts in total, where each rank holds $E/R$ experts with disjoint expert sets across ranks. Let $\mathcal{E}_r$ denote the set of experts held by rank $r$, and let $r_e$ denote the home rank of expert $e$. For a given micro-batch, let $n_{re}$ denote the number of tokens sent by rank $r$ to expert $e$. After All-to-All dispatch, the actual computational load on rank $r$ is
\begin{equation}
\ell_{r} = \sum_{e \in \mathcal{E}_r} \sum_{r'=0}^{R-1} n_{r'e}.
\end{equation}
Let the global mean load be $\bar{\ell} = \frac{1}{R}\sum_{r} \ell_r$. The excess load and spare capacity of rank $r$ are
\begin{equation}
\mathrm{excess}_r = \max(\ell_r - \bar{\ell},\ 0), \quad \mathrm{spare}_r = \max(\bar{\ell} - \ell_r,\ 0).
\end{equation}
The load imbalance ratio is
\begin{equation}
\rho = \frac{\max(\ell_0,\, \ell_1,\, \ldots,\, \ell_{R-1}) - \bar{\ell}}{\bar{\ell}}.
\end{equation}
In synchronous training, overall latency is determined by $\max_r \ell_r$; reducing $\rho$ to near zero is the primary objective. The sets of hot ranks and cold ranks are
\begin{equation}
\mathcal{H} = \{r \mid \mathrm{excess}_r > 0\}, \quad \mathcal{C} = \{r \mid \mathrm{spare}_r > 0\}.
\end{equation}

Let $z_{er} \in \{0,1\}$ be the replica placement indicator, where $e$ is an expert held by a hot rank ($r_e \in \mathcal{H}$) and $r \in \mathcal{C}$ is the target cold rank. Once a placement scheme $\mathbf{z}$ is determined, token routing is adjusted accordingly; let $\rho(\mathbf{z})$ denote the residual imbalance after adjustment. The optimization objective is
\begin{equation}
\begin{aligned}
\min_{\mathbf{z} \in \{0,1\}^{E \times |\mathcal{C}|}} & \quad \underbrace{\rho(\mathbf{z})}_{\text{residual imbalance}} + \mu \cdot \underbrace{\sum_{e,r} z_{er} \cdot W_{r_e,r}}_{\text{weighted replica comm.\ cost}} \\
\text{s.t.} & \quad \sum_{e} z_{er} \leq K, \quad \forall r \in \mathcal{C}
\end{aligned}
\end{equation}
where $\mu > 0$ is a trade-off coefficient between balancing effectiveness and communication cost, and $W_{r_e,r}$ is the communication cost from the home rank $r_e$ of expert $e$ to target cold rank $r$, defined by the cluster topology:
\begin{equation}
W_{r_e,r} = \begin{cases}
1      & \text{if } r_e \text{ and } r \text{ are on the same node} \\
\lambda & \text{if } r_e \text{ and } r \text{ are on different nodes.}
\end{cases}
\end{equation}
The parameter $\lambda > 1$ reflects the bandwidth disparity between inter-node and intra-node links; we set $\lambda = 3$ in our experiments. The fundamental limitation of existing replica placement strategies is that their objectives include only the $\rho(\mathbf{z})$ term, with communication cost $W_{r_e,r}$ entirely absent, causing replicas to scatter across inter-node ranks and incurring unnecessary inter-node communication overhead. This paper incorporates $W_{r_e,r}$ into the replica-placement objective, weighted by the 0/1 indicator $z_{er}$, to reflect the fixed topology-dependent cost of transmitting guest expert weights during expert dispatch. Because this objective jointly involves discrete replica selection, spare-slot capacity constraints, and integer token assignment, we do not directly solve the mixed-integer problem. Instead, we use Sinkhorn-Knopp iterations to construct a rank-level topology-aware flow hint, and then generate an executable plan through discrete matching and token assignment.

\subsection{A.2\quad Phase 1: Sinkhorn-Knopp Topology-Aware Flow Planning}

The bi-objective problem above involves both discrete replica placement decisions and continuous token allocation, and direct joint solving causes the problem scale to grow rapidly with the EP degree. We decompose it into three cooperative, progressively refined phases. Phase~1 uses balanced optimal transport to generate rank-level topology-aware soft flow hints; Phase~2 combines the flow hints with residual spillover and spare capacity to produce integer replica assignments at expert granularity, using column-priority competition to ensure fair scheduling across ranks; Phase~3 performs precise token-to-spare-slot assignment at token granularity via a Lagrangian auction.

When deciding which hot ranks' excess load should be offloaded to which cold ranks, we prefer routing overflow along intra-node bandwidth. This is a global trade-off problem in which greedy incremental allocation cannot guarantee globally consistent topology preferences. Phase~1 formulates a rank-level topology-cost optimal transport problem: the excess-load vector $\mathbf{s} \in \mathbb{R}^{R}$ provides supply, and the spare-capacity vector $\mathbf{d} \in \mathbb{R}^{R}$ provides demand, with zero supply on non-hot ranks and zero demand on non-cold ranks. Since both vectors are defined with respect to the same mean load $\bar{\ell}$, they have equal total mass in exact arithmetic:
\begin{equation}
\sum_{r=0}^{R-1} s_r = \sum_{r=0}^{R-1} d_r.
\end{equation}
Therefore, Phase~1 uses balanced OT rather than unbalanced OT. Given the topology cost matrix $W \in \mathbb{R}^{R\times R}$, the rank-level transport plan $T^* \in \mathbb{R}^{R \times R}_{\geq 0}$ satisfies equality marginal constraints:
\begin{equation}
\begin{aligned}
T^* = \arg\min_{T \geq 0} & \quad \langle T, W \rangle \\
\text{s.t.} & \quad T\mathbf{1} = \mathbf{s},\quad T^\top\mathbf{1} = \mathbf{d}.
\end{aligned}
\end{equation}
The raw OT problem requires LP solving, which is complex to implement on GPU. Introducing a negative-entropy regularization term $H(T) = -\sum_{ij} T_{ij} \log T_{ij}$ relaxes it to
\begin{equation}
\begin{aligned}
T^*_\varepsilon = \arg\min_{T \geq 0} & \quad \langle T, W \rangle - \varepsilon H(T) \\
\text{s.t.} & \quad T\mathbf{1} = \mathbf{s},\quad T^\top\mathbf{1} = \mathbf{d}.
\end{aligned}
\end{equation}
The optimal solution of this relaxed problem possesses a Gibbs kernel structure. Constructing the Gibbs kernel matrix $M \in \mathbb{R}^{R \times R}$ with elements $M_{ij} = \exp(-W_{ij}/\varepsilon)$, the optimal solution takes the form
\begin{equation}
T^*_\varepsilon = \mathrm{diag}(\mathbf{u})\, M\, \mathrm{diag}(\mathbf{v}).
\end{equation}
Setting $\varepsilon = \lambda$, the kernel value ratio between intra-node and inter-node entries is
\begin{equation}
\frac{M_{\mathrm{intra}}}{M_{\mathrm{inter}}} = \exp\!\left(\frac{\lambda-1}{\lambda}\right).
\end{equation}
This induces soft topological preferences. Unlike LPLB, which imposes predefined graph topologies (Cube, Torus) as hard constraints, soft preferences preserve a larger candidate space under extreme imbalance: when intra-node capacity is insufficient, inter-node ranks remain feasible candidates with higher cost rather than being removed by a fixed graph. The scaling vectors $(\mathbf{u}, \mathbf{v})$ are solved via alternating Sinkhorn-Knopp GEMV iterations:
\begin{equation}
\mathbf{u}^{(t+1)} = \frac{\tilde{\mathbf{s}}}{M\,\mathbf{v}^{(t)}}, \qquad
\mathbf{v}^{(t+1)} = \frac{\mathbf{d}}{M^\top \mathbf{u}^{(t+1)}},
\end{equation}
where $\tilde{\mathbf{s}}=\gamma\mathbf{s}$ and $\gamma=\min(1,\sum_r d_r / \max(\sum_r s_r,\epsilon_0))$. In exact arithmetic, $\tilde{\mathbf{s}}=\mathbf{s}$; this scaling is only an implementation safeguard against integer mean loads, finite precision, or anomalous inputs that make the total supply slightly exceed the total demand. After a fixed number of iterations, we obtain $T=\mathrm{diag}(\mathbf{u})M\mathrm{diag}(\mathbf{v})$. The implementation further applies a column-cap safeguard:
\begin{equation}
T_{ij}\assign{}T_{ij}\cdot \min\!\left(1,\frac{d_j}{\sum_i T_{ij}+\epsilon_0}\right),
\end{equation}
which ensures that the flow hint does not numerically exceed each target rank's spare capacity. These supply scaling and column cap operations are not an unbalanced OT solver; they are feasibility safeguards around a balanced Sinkhorn flow hint. The Phase~1 output is not executed directly. Instead, the rank-level plan is expanded to the expert dimension according to each expert's home rank. This yields the OT flow hint matrix $T_{\mathrm{er}} \in \mathbb{R}^{E \times R}$, where each element $(T_{\mathrm{er}})_{e,r}=T_{r_e,r}$ represents the Phase~1 flow strength for offloading expert $e$ from its home rank $r_e$ to cold rank $r$. Its normalized form is used as the third-level scoring signal in Phase~2.

\subsection{A.3\quad Phase 2: Column-Priority Iterative Matching}

Phase~1 provides a rank-level flow hint, but replica placement is fundamentally an integer decision. Each cold rank accommodates at most $K$ complete expert replicas, and multiple cold ranks may simultaneously contend for the same hotspot expert. Phase~2 combines this hint with residual spillover and spare capacity to produce concrete binary decisions of the form which expert is placed in which spare slot.

Within each hot rank, experts are sorted in ascending order of load, and the spillover amount $\mathrm{spill}_e$ that each expert can externally absorb is computed via cumulative excess differencing. Let $x_{r,1} \leq x_{r,2} \leq \cdots \leq x_{r,E/R}$ be the sorted expert loads; the cumulative excess of rank $r$ at position $k$ is
\begin{equation}
\sigma_{r,k} = \max\!\left(\sum_{j=1}^{k} x_{r,j} - \bar{\ell},\ 0\right),
\end{equation}
and the marginal spillover contribution at sorted position $k$ is
\begin{equation}
\tilde{s}_{r,k} = \sigma_{r,k} - \sigma_{r,k-1}, \quad (\sigma_{r,0} = 0).
\end{equation}
After remapping to original expert indices we obtain $\mathrm{spill} \in \mathbb{R}^E$, ensuring lightly loaded experts have zero spillover while heavily loaded experts progressively assume spill responsibility in proportion to their marginal contribution. Meanwhile, $\mathrm{spare}_r=\max(\bar{\ell}-\ell_r,0)$ denotes the remaining capacity that cold rank $r$ can accept relative to the mean load $\bar{\ell}$. Therefore, when expert $e$ is placed on cold rank $r$, at most $\min(\mathrm{spill}_e,\mathrm{spare}_r)$ tokens can be redistributed, which represents the maximum load-balance improvement achievable by this match.

Phase~2 selects $K$ hot experts for each cold rank $r$ and determines allocation amounts $g_{er} \geq 0$ to maximize total redistributed load:
\begin{equation}
\begin{aligned}
\max_{g_{er} \geq 0} & \quad \sum_{e,\, r \in \mathcal{C}} g_{er} \\
\text{s.t.} & \quad \sum_{r} g_{er} \leq \mathrm{spill}_e, \quad \forall e \\
            & \quad \sum_{e} g_{er} \leq \mathrm{spare}_r, \quad \forall r \in \mathcal{C} \\
            & \quad |\{e \mid g_{er}>0\}| \leq K, \quad \forall r \in \mathcal{C}.
\end{aligned}
\end{equation}
A three-level priority score is defined for each (cold rank $r$, hot expert $e$) pair:
\begin{equation}
\begin{split}
\mathrm{score}_{er} = &\underbrace{\min(\mathrm{spill}_e,\ \mathrm{spare}_r)}_{\text{primary: balance gain}} \\
&+ \alpha \underbrace{B_{er}}_{\substack{\text{secondary:}\\\text{topo.\ pref.}}} + 0.1\alpha \underbrace{(T_{\mathrm{er}})_{\mathrm{norm}}}_{\substack{\text{tertiary:}\\\text{OT hint}}}
\end{split}
\end{equation}
where $\alpha = \bar{\ell} R / E$ is a dimensional alignment coefficient ensuring the primary term dominates the secondary, so balance effectiveness takes priority over topological preference. $(T_{\mathrm{er}})_{\mathrm{norm}}=T_{\mathrm{er}}/\max(\mathrm{excess}_{r_e},\epsilon_0)$ denotes the normalized OT flow hint. It measures the relative strength of the Phase~1 recommendation to offload expert $e$ from its home rank $r_e$ to cold rank $r$, and serves as a global-optimality tiebreaker. The topology preference matrix $B_{er}$ is
\begin{equation}
B_{er} = \begin{cases}
1              & \text{if } e \text{ and rank } r \text{ are intra-node} \\
w_{\mathrm{inter}} & \text{if } e \text{ and rank } r \text{ are inter-node, } w_{\mathrm{inter}} < 1 \\
0              & \text{if } e \text{ belongs to rank } r \text{ (home).}
\end{cases}
\end{equation}
The choice of matching direction is critical. If experts are iterated as the outer loop (row-priority), heavily loaded experts greedily occupy all spare slots, leaving lightly loaded ranks with insufficient scheduling opportunities and residual imbalance of 7--10\%. To address this, we propose column-priority iterative matching from the cold rank perspective. In each round, every cold rank independently selects its highest-scoring candidate expert. If multiple cold ranks select the same expert, a conflict arbitration mechanism keeps the best match and lets the losing cold ranks re-enter the next round. This perspective shift guarantees each cold rank an independent competition opportunity per round, reducing residual imbalance to 1--2\%. The detailed procedure is given in Algorithm~\ref{alg:phase2}; the outer loop runs $K$ times, the inner loop resolves conflicts with a fixed iteration count, and the entire procedure is free of dynamic branches.

\begin{algorithm}[t]
  \caption{Column-Priority Iterative Matching (Phase~2)}
  \label{alg:phase2}
  \begin{algorithmic}[1]
    \REQUIRE Spillover $\mathrm{spill}_e$, spare capacity $\mathrm{spare}_r$, topology preference matrix $B_{er}$, OT hint $T_{\mathrm{er}}$, replica count $K$
    \ENSURE $A \in \mathbb{Z}_{\geq 0}^{E \times S}$: planned token count from expert $e$ to spare slot $s$ ($S = |\mathcal{C}| \times K$)
    \STATE $U_{er}\assign0\ (\forall e,r)$;\quad $A_{es}\assign0\ (\forall e,s)$ \COMMENT{$U_{er}{=}1$: expert $e$ assigned to rank $r$}
    \FOR{$k\assign0, 1, \ldots, K-1$}
      \STATE $m_r\assign0\ (\forall r)$ \COMMENT{cold ranks matched in this round}
      \FOR{$\mathrm{inner}\assign0, 1, \ldots, |\mathcal{C}|-1$}
        \STATE \textit{// Step 1: Compute scores and mask invalid entries}
        \STATE $\mathrm{score}_{er}\assign{}\min(\mathrm{spill}_e,\mathrm{spare}_r)\,+\,\alpha B_{er}\,+\,0.1\alpha(T_{\mathrm{er}})_{\mathrm{norm}}$
        \IF{$e \in \mathcal{E}_r$ \textbf{or} $U_{er}{=}1$ \textbf{or} $m_r{=}1$ \textbf{or} $\mathrm{spare}_r{\leq}0$ \textbf{or} $\mathrm{spill}_e{\leq}0$}
          \STATE $\mathrm{score}_{er}\assign-\infty$
        \ENDIF
        \STATE \textit{// Step 2: Cold ranks choose best experts}
        \STATE $e^*_r\assign{}\arg\max_e\;\mathrm{score}_{er}$;\; $\mathcal{V}\assign{}\{r\,|\,\max_e\,\mathrm{score}_{er}{>}-\infty\}$
        \STATE \textit{// Step 3: Resolve expert-rank conflicts}
        \STATE $P_{er}\assign\mathbf{1}[r \in \mathcal{V}]\cdot\mathbf{1}[e^*_r{=}e]$
        \STATE $r^*_e\assign\arg\max_r\,(\mathrm{score} \odot P)_{er}$\quad $(\forall\, e\!:\!\sum_r P_{er}{>}0)$
        \STATE $\delta_{er}\assign\mathbf{1}[r^*_e{=}r]\cdot P_{er}$
        \STATE \textit{// Step 4: Update assignments and residuals}
        \STATE $g_{er}\assign\delta_{er}\cdot\min(\mathrm{spill}_e,\;\mathrm{spare}_r)$
        \STATE $\mathrm{spill}_e \mathrel{-}= \textstyle\sum_r g_{er}$;\quad $\mathrm{spare}_r \mathrel{-}= \textstyle\sum_e g_{er}$
        \STATE $U_{er}\assign{}U_{er} \vee \delta_{er}$;\quad $m_r\assign{}m_r \vee (\textstyle\sum_e \delta_{er}{>}0)$
      \ENDFOR
      \STATE $A\assign{}A + \mathrm{scatter}(g,\;\text{spare slot }k)$
    \ENDFOR
  \end{algorithmic}
\end{algorithm}

\subsection{A.4\quad Phase 3: Lagrangian Auction Token Assignment}

Phase~2 determines that expert $e$ plans to assign $A_{es}$ tokens to spare slot $s$; however, tokens for expert $e$ are distributed across multiple ranks ($n_{r'e}$ tokens per rank $r'$), requiring determination of the actual contribution $X_{r'es}$ from each rank $r'$, subject to
\begin{equation}
\sum_{r'} X_{r'es} = A_{es}, \quad \sum_{s} X_{r'es} \leq n_{r'e}, \quad X_{r'es} \geq 0.
\end{equation}
Proportional allocation introduces floating-point rounding errors and ignores topological preference. Phase~3 therefore introduces a Lagrangian auction mechanism. Introducing Lagrange multipliers (prices) $p_{r'} \geq 0$ for the rank capacity constraints, the augmented objective is
\begin{equation}
\begin{aligned}
\max_{X \geq 0} & \quad \sum_{r', s} \left( B_{r's} - p_{r'} \right) X_{r's} \\
\text{s.t.} & \quad \sum_{r'} X_{r's} \leq d_s,\quad X_{r's} \leq n_{r'e_s},
\end{aligned}
\end{equation}
where $B_{r's}$ is the topology bonus ($1$ for intra-node, $w_{\mathrm{inter}}$ for inter-node) and $d_s = A_{e_s s}$ is the planned allocation for spare slot $s$. In each round, each spare slot bids for a source rank by net profit:
\begin{equation}
\mathrm{winner}(s) = \arg\max_{r'}\, (B_{r's} - p_{r'}).
\end{equation}
After a winning rank completes its allocation, its price monotonically increases ($p_{r'}\assign{}p_{r'} + \varepsilon$), naturally reducing its competitiveness in subsequent rounds, thereby achieving fair load distribution with embedded topological preference. A fixed iteration count ensures CUDA Graph compatibility, eliminates floating-point truncation, and guarantees exact token conservation. The complete procedure is given in Algorithm~\ref{alg:phase3}.

\begin{algorithm}[t]
  \caption{Lagrangian Auction Token Assignment (Phase~3)}
  \label{alg:phase3}
  \begin{algorithmic}[1]
    \REQUIRE Token counts $n_{re}$ (per rank per expert), planned allocation $A_{es}$ (from Alg.~\ref{alg:phase2}), topology bonus $B_{rs}$, price step $\varepsilon$, iteration count $T_{\max}$
    \ENSURE $X_{rs} \in \mathbb{Z}_{\geq 0}$: actual token count from rank $r$ to spare slot $s$
    \STATE $p_r\assign0\ (\forall r)$;\quad $X_{rs}\assign0\ (\forall r,s)$ \COMMENT{Initialize Lagrange multipliers}
    \FOR{$t\assign1, 2, \ldots, T_{\max}$}
      \STATE \textit{// Step 1: Spare slots bid for source ranks}
      \STATE $\mathrm{score}_{rs}\assign{}B_{rs} - p_r$
      \IF{$n_{r,e_s} \leq 0$ \textbf{or} $A_{e_s,s} \leq 0$}
        \STATE $\mathrm{score}_{rs}\assign-\infty$
      \ENDIF
      \STATE $r^*_s\assign\arg\max_r\;\mathrm{score}_{rs}$
      \STATE \textit{// Step 2: Ranks serve highest-profit slots}
      \STATE $c_{rs}\assign\mathbf{1}[r^*_s{=}r]$
      \STATE $s^*_r\assign\arg\max_s\,(\mathrm{score} \odot c)_{rs}$
      \STATE \textit{// Step 3: Compute allocation (bounded by $n_{re}$, $A_{es}$)}
      \STATE $x_r\assign{}\min(n_{r,e_{s^*_r}},\;A_{e_{s^*_r},s^*_r})\cdot\mathbf{1}[s^*_r \text{ valid}]$
      \STATE \textit{// Step 4: Update $X$, $n$, $A$, and Lagrange multipliers}
      \STATE $X_{r,s^*_r}\assign{}X_{r,s^*_r} + x_r$
      \STATE $n_{r,e_{s^*_r}}\assign{}n_{r,e_{s^*_r}} - x_r$
      \STATE $A_{e_{s^*_r},s^*_r}\assign{}A_{e_{s^*_r},s^*_r} - x_r$
      \STATE $p_r\assign{}p_r + \varepsilon\cdot\mathbf{1}[x_r{>}0]$
    \ENDFOR
  \end{algorithmic}
\end{algorithm}

\section{B\quad Supplementary Details for the Experiments}

This appendix reports the full computing environment (\S B.1), the complete set of training hyperparameters (\S B.2), and the implementation and usage of TAOT (\S B.3).

\subsection{B.1\quad Computing Environment}

All end-to-end experiments run on four training nodes, each with eight NVIDIA A800-SXM4-80GB GPUs (32 GPUs in total). GPUs within a node are connected by NVLink, and inter-node communication uses InfiniBand; this two-tier interconnect is exactly the source of the intra-/inter-node cost gap modeled by the topology cost matrix $W$, with the inter-node factor set to $\lambda=3$ in our experiments. The distributed training system is an in-house framework built on Megatron-Core. Table~\ref{tab:env} lists the detailed hardware and software configuration of each node.

\begin{table}[h]
\centering
\small
\setlength{\tabcolsep}{5pt}
\renewcommand{\arraystretch}{1.15}
\begin{tabular}{@{}ll@{}}
\toprule
\textbf{Component} & \textbf{Specification} \\
\midrule
CPU        & 2$\times$ Intel Xeon Platinum 8350C @ 2.60\,GHz \\
           & (32 cores/socket, 2 threads/core) \\
Memory     & 1.5\,TiB DRAM \\
GPU        & 8$\times$ NVIDIA A800-SXM4-80GB \\
Intra-node & NVLink \\
Inter-node & InfiniBand \\
OS         & Ubuntu 24.04.2 LTS (kernel 5.15.0-124) \\
GPU driver & 535.230.02 \\
CUDA       & 12.9 (V12.9.86) \\
PyTorch    & 2.8.0 (NVIDIA NGC 25.06) \\
cuDNN / NCCL & 9.16.0 / 2.27.3 \\
Libraries  & transformers 5.3.0, triton 3.3.0 \\
\bottomrule
\end{tabular}
\caption{Per-node hardware and software environment.}
\label{tab:env}
\end{table}

\subsection{B.2\quad Training Hyperparameters}

The end-to-end evaluation uses the Qwen3-30B-A3B MoE model on the Pile-test dataset under the TP4\,PP2\,EP16 parallel configuration. Since TAOT operates purely at the system-execution level and does not alter the routing semantics or the model outputs, its balancing and communication behavior is independent of the specific corpus; Pile-test only serves as a representative pretraining workload that produces realistic expert-routing traffic. To prevent routing-level regularization from masking the behavior of system-level dynamic balancing, the MoE auxiliary load-balancing loss is disabled, so the routing distribution is determined by the model itself. Table~\ref{tab:hparams} lists the complete set of hyperparameters.

\begin{table}[h]
\centering
\small
\setlength{\tabcolsep}{6pt}
\renewcommand{\arraystretch}{1.15}
\begin{tabular}{ll}
\toprule
\textbf{Hyperparameter} & \textbf{Value} \\
\midrule
\multicolumn{2}{l}{\textit{Model / sequence}} \\
Model              & Qwen3-30B-A3B (MoE) \\
Rotary base        & 1{,}000{,}000 \\
Sequence length    & 32{,}768 \\
Max position emb.  & 32{,}768 \\
Init method std    & 0.006 \\
\midrule
\multicolumn{2}{l}{\textit{Optimization}} \\
Micro-batch size   & 3 \\
Global batch size  & 96 \\
Optimizer          & Adam ($\beta_1{=}0.9$, $\beta_2{=}0.95$, $\epsilon{=}10^{-8}$) \\
Learning rate      & $1.0\times10^{-5}$ (min $1.0\times10^{-6}$) \\
LR schedule        & cosine, warmup fraction 0.002 \\
Train / decay iters & 50{,}000 / 50{,}000 \\
Weight decay       & 0.1 \\
Grad clip          & 1.0 \\
Norm epsilon       & $10^{-6}$ \\
Precision          & bf16 (precision-aware optimizer) \\
Initial loss scale & 65{,}536 \\
\midrule
\multicolumn{2}{l}{\textit{MoE / routing}} \\
Router top-$k$     & 8 \\
Router dtype       & fp32 \\
Aux-loss           & disabled \\
Grouped GEMM       & enabled \\
\midrule
\multicolumn{2}{l}{\textit{Parallelism / memory}} \\
TP / PP / EP       & 4 / 2 / 16 \\
Expert TP size     & 1 \\
Virtual PP stages  & 3 \\
Sequence parallel  & enabled \\
Distributed optimizer & ZeRO-1 \\
Recomputation      & full, block, 24 layers \\
Comm.\ backend     & NCCL \\
\bottomrule
\end{tabular}
\caption{End-to-end training hyperparameters.}
\label{tab:hparams}
\end{table}

\subsection{B.3\quad Implementation and Usage}

TAOT is integrated into the training framework as an extension of the guest-expert mechanism and is enabled through command-line arguments. Two functions are exposed to the user. The topology-aware expert-dispatch planner is selected with \texttt{-{}-moe-echo-algorithm}, which chooses the specific planning algorithm used in the planning phases. The communication overlap is turned on with \texttt{-{}-moe-echo-expert-dispatch-overlap}, which overlaps guest-expert weight transfer with home-expert computation to hide the communication latency. A minimal configuration enables \texttt{-{}-moe-enable-echo}, sets the number of spare (guest) expert slots per rank with \texttt{-{}-moe-num-echo-experts}, and selects the planning algorithm with \texttt{-{}-moe-echo-algorithm sinkhorn} (\texttt{sinkhorn} for TAOT). Table~\ref{tab:flags} lists all available command-line arguments.

\begin{table*}[t]
\centering
\small
\setlength{\tabcolsep}{8pt}
\renewcommand{\arraystretch}{1.2}
\begin{tabular}{@{}ll@{}}
\toprule
\textbf{Argument} & \textbf{Description} \\
\midrule
\multicolumn{2}{@{}l}{\textit{Core}} \\
\texttt{-{}-moe-enable-echo} & Enable guest-expert load balancing. \\
\texttt{-{}-moe-num-echo-experts K} & Spare expert slots per rank ($K$; a multiple of the EP degree). \\
\texttt{-{}-moe-echo-algorithm sinkhorn} & Planning algorithm: \texttt{sinkhorn} (TAOT). \\
\midrule
\multicolumn{2}{@{}l}{\textit{Efficiency}} \\
\texttt{-{}-moe-echo-expert-dispatch-overlap} & Overlap guest-expert dispatch with home-expert computation. \\
\texttt{-{}-moe-echo-expert-dispatcher-type} & Collective backend for expert-parameter fetch: \texttt{alltoall} or \texttt{hybridep}. \\
\midrule
\multicolumn{2}{@{}l}{\textit{Diagnostics}} \\
\texttt{-{}-moe-echo-log-steps 1,3,5} & Log dispatch statistics at the given global steps. \\
\texttt{-{}-moe-echo-log-layers 1,2,3,4} & Restrict logging to the given transformer layers. \\
\texttt{-{}-moe-echo-log-file PATH} & Output file for dispatch diagnostics. \\
\bottomrule
\end{tabular}
\caption{Command-line arguments exposed by TAOT.}
\label{tab:flags}
\end{table*}

A few arguments deserve further explanation. \texttt{-{}-moe-num-echo-experts} sets the number of spare (guest) expert slots reserved per rank ($K$); it must be a multiple of the EP degree so that the reserved slots are distributed evenly across all EP ranks. \texttt{-{}-moe-echo-expert-dispatcher-type} selects the communication backend used to fetch guest-expert parameters: \texttt{alltoall} uses a plain All-to-All, while \texttt{hybridep} uses a dedicated communication library as an alternative backend. 

The three diagnostic arguments control logging. \texttt{-{}-moe-echo-log-steps} specifies the global training steps at which dispatch statistics are recorded (e.g.\ \texttt{1,3,5}); \texttt{-{}-moe-echo-log-layers} restricts logging to the given transformer layers (e.g.\ \texttt{1,2,3,4}); and \texttt{-{}-moe-echo-log-file} sets the output file path for the diagnostics. When enabled, TAOT produces a per-step, per-layer report organized into four parts: (1) the per-rank and per-expert load before echo routing, with hot ranks and hot experts marked; (2) the cloning plan, listing for each cloned hot expert its target spare slot, whether the copy is intra- or inter-node, and the number of tokens rerouted, followed by a topology summary of intra-/inter-node copy counts and the weighted communication cost; (3) the projected per-rank load after routing; and (4) a load-balance improvement summary reporting the imbalance before and after, the peak-load reduction, the fraction of tokens rerouted, and the final intra-/inter-node copy counts and weighted cost. This report makes the balancing behavior of each planning invocation directly inspectable.

\end{document}